\input harvmac

\vskip 1cm

 \Title{ \vbox{\baselineskip12pt\hbox{  Brown Het-1191 }}}
 {\vbox{
\centerline{  Gravity from  CFT on $S^N ( X )$:    }
\centerline{    Symmetries and Interactions   }  }}

\centerline{$\quad$ { Antal Jevicki, 
        Mihail Mihailescu and  Sanjaye Ramgoolam }}
\smallskip
\centerline{{\sl  }}
\centerline{{\sl Brown  University}}
\centerline{{\sl Providence, RI 02912 }}
\centerline{{\tt antal, mm, ramgosk@het.brown.edu}}
 \vskip .3in 
 
 The orbifold CFT dual to 
 string theory  on  $ADS_3 \times S^3$ 
 allows a construction of gravitational actions 
 based on collective field techniques. We describe a fundamental
 role  played 
 by a Lie algebra constructed 
 from chiral primaries and their CFT conjugates. 
 The leading terms in the algebra at large $N$ are derived 
 from the computation of  chiral primary
 correlation functions. The algebra is argued to determine the 
 dynamics of the  theory, its representations provide
 free and interacting hamiltonians for chiral primaries.
 This dynamics  is seen to be given
 by an effective one plus one dimensional field theory.
 The structure of the algebra and its representations 
 shows qualitatively new features associated with 
 thresholds at  $L_0= N$, $ L_0 = N/2$ and  $L_0=N/4$, 
 which are related to the stringy exclusion principle 
 and to black holes.  We observe relations  
 between fusion rules of $SU_q(2|1,1)$ for $q = e^{i \pi \over {N+1}} $,
 and the correlation functions, which provide further 
 evidence for a non-commutative spacetime.


\Date{11/98}

\lref\aj{J.Avan and A.Jevicki, Phys.Lett.B266:35,1991 ;Phys.Lett.
B272:17-24,1991 }
\lref\vafa{C.Vafa ,{ \it Puzzles at Large N }, hep-th/9804172 }
\lref\moorebps{J.H.Harvey and G. Moore, 
{\it On the algebras of BPS states,}  hepth/9609017. } 
\lref\cds{ A. Connes, M. Douglas and A. Schwarz, { \it Non-commutative 
            Geometry and Matrix Theory : Compactification on tori.} 
           JHEP 02 (1998) 003.  } 
\lref\malda{ J. Maldacena, { \it  The large N limit of superconformal 
                             field theories and supergravity,} 
                      Adv.Theor.Math.Phys.2: 231-252, 1998, 
                       hepth/9711200 } 
\lref\witten{ E. Witten, 
{\it Anti-de-Sitter space and holography,} hepth-9802150 } 
\lref\gkp {S.S.Gubser,I.R.Klebanov and A.M.Polyakov, 
   {\it Gauge Theory Corellators from Non-Critical String Theory,}
     hepth-9802109 }
\lref\jr{  A. Jevicki, S. Ramgoolam, 
{ \it Non commutative gravity from the ADS/CFT
          correspondence,}  JHEP 9904 (1999) 032,  hep-th/9902059 } 
\lref\malstro{  J. Maldacena, 
A. Strominger,{ \it AdS3 Black Holes and a Stringy
           Exclusion Principle,}  JHEP 9812 (1998) 005, hep-th/9804085 } 
\lref\Ver{ R. Dijkgraaf, G. Moore, 
E. Verlinde, H. Verlinde, { \it Elliptic Genera
          of Symmetric Products and Second Quantized Strings,} 
          Commun.Math.Phys. 185 (1997) 197-209, hep-th/9608096 } 
\lref\Frol{ G.E.Arutyunov, S.A.Frolov, { \it 
 Four graviton scattering amplitude
           from $S^N{{\bf R}}^{8}$ supersymmetric orbifold sigma model,} 
           Nucl.Phys. B524 (1998) 159-206, hep-th/9712061 } 
\lref\rsal{  L. Rozansky and H. Saleur, Nucl. Phys. B 389 (1993) 365 }
\lref\Keller{ G. Keller, {\it Fusion Rules of $U_qSL(2,C)$, $q^m = 1$,}
                Lett. Math. Phys. 21: 273, 1991 } 
\lref\db{J. de Boer,{ \it Large N Elliptic Genus 
and AdS/CFT Correspondence }, JHEP 9905 (1999) 017, hep-th/9812240 }
\lref\seiberg{S.Lee,S.Minwalla,M.Rangamani and N.Seiberg,
{ \it Three point functions of Chiral operators in $D=4$, 
   $N=4$ SYM at lareg $N$ } hep-th/9806074. }
\lref\mathur{D.Freedman,S.Mathur,A.Mathusis and L.Rastelli,hep-th/9804058. }
\lref\zucchini{F.Bastianelli and R.Zucchini,hep-th/9907047.}
\lref\vp{Vipul Periwal, { \it String Field Theory Hamiltonians 
 from Yang-Mills Theories,}  hep-th/9906052. }
\lref\ml{F. Larsen, E. Martinec, { \it U(1) Charges and Moduli in the D1-D5
         System, JHEP 9906 (1999) 019,}  hep-th/9905064. } 
\lref\bhdm{T. Banks, M. Douglas, G. Horowitz, E. Martinec, 
{ \it AdS dynamics from conformal Field Theory,} hepth/9808016.}
\lref\dj{S. R. Das, A. Jevicki, String field theory and physical  
         interpretation of D=1 strings, Mod.Phys.Lett.A5:1639-1650,1990. }
\lref\js{A. Jevicki and B.Sakita ,Nucl.Phys.B165:511,1980;
        Nucl.Phys.B185:89,1981 }
\lref\lvw{ W. Lerche, C. Vafa, N. P. Warner,  Chiral rings in
          N=2 superconformal theories,  Nucl.Phys.B324:427,1989 }
\lref\ber{ F.A. Berezin, {\it Models of Gross-Neveu 
          type are quantization of a classical Mechanics with non-linear
          phase space,} 
  Commun. in Math. Physics, 63, 131-153 (1978);
  A.Jevicki and N.Papanicolaou,
  { \it Classical Dynamics at Large N } ,Nucl.Phys.B171:362,1980, } 
\lref\reshtu{ N. Reshetikhin, V.G. Turaev, 
{ \it Invariants of three manifolds  via 
link polynomials and quantum groups,}  Invent.Math.103:547-597,1991}
\lref\gepwit{D. Gepner, E. Witten, { \it String theory on group manifolds, } 
             Nucl.Phys.B278:493,1986}  
\lref\mathint{P. Goddard, D. Olive, { \it Kac-Moody and Virasoro Algebras 
                                     in relation to Quantum Physics,} 
                       IJMPA, vol.1, No. 2, (1986) 303.  }
\lref\ggm{ Per Berglund, Eric. Gimon, Djordje Minic, 
             { \it The AdS/CFT correspondence and Spectrum 
             generating algebras,}  hepth-9905097. }  
\lref\mms{ J.Maldacena, G.Moore, A.Strominger,
{ \it Counting BPS black holes in Toroidal Type II string theory,} 
hep-th/9903163. } 
\lref\Sugaw{ Y. Sugawara, { \it  N=(0,4) Quiver $SCFT_2$ and Supergravity
            on $AdS_3 \times S^2$ } ,JHEP 9906 (1999) 035, hep-th/9903120 }
\lref\tesh{J. Teschner, { \it The deformed two-dimensional black hole,} 
    hepth/9902189.  } 
\lref\chk{ C.S.Chu, P.M.Ho, Y.C.Kao, {\it World-volume Uncertainty \
                  relations for D-branes,} ~ hepth/9904133. }   
\lref\bv{ M. Berkooz and H. Verlinde, {\it Matrix Theory, AdS/CFT and
Higgs-Coulomb Equivalence,}  hep-th/9907100. } 
\lref\madore{J. Madore, { \it Non-commutative Geometry for Pedestrians,} 
grqc/9906059. } 
\lref\dfms{ L. Dixon, D. Friedan, E. Martinec and S. Shenker, 
            { \it The conformal field theory of orbifolds,} 
            Nucl. Phys. B282 (1987) 13. } 
\lref\jky{ A. Jevicki, Y. Kazama, T. Yoneya, 
{ \it Generalized conformal symmetry in D-brane Matrix models } 
 Phys. Rev. D59, 1999, hepth/9810146. } 
\lref\sw{ N. Seiberg, E. Witten, { \it The D1/D5 system and singular CFT,}
   hepth/9903224.   } 
\lref\mpy{ D. Minic, J. Polchinski, Z. Yang, 
 {\it Translation invariant backgrounds in $1 +1$ dimensional string theory } 
 Nucl. Phys. B369, 324, 1992 }  
\lref\jv{ Antal Jevicki and Andre van Tonder, ``Finite [ q-oscillator ] 
                                   Description of 2D string theory,''
Mod.Phys.Lett.A11:1397-1410,(1996),hep-th/9601058. }

\newsec{ Introduction }
In the context of the ADS/CFT duality \refs{ \malda, \gkp, \witten } 
we began in  \jr\ 
(hereafter referred to as I ) a construction
of elements of supergravity on $ADS_3 \times S^3$ 
 based on a simple orbifold conformal field 
theory with target space $S^N(T^4)$ or $S^N(K3)$. The novel 
feature of the emerging gravitational theory is a 
`stringy  exclusion principle'
(\malstro ) which follows from the  CFT. 

We argued that this exclusion principle 
implies  a role for a non-commutative  
spacetime.
Similar aspects of non-commutativity 
in  closed string backgrounds were also found recently
in \bv.
The precise  form of non-commutative 
spacetime suggested in I was  
$SU(2)_q \times SU_q(1,1 )$
with $q= e^{ i \pi \over {N+1}}$. 
One way to arrive at the value of $q$ 
 was geometrical. A deformation of 
the $S^3$ sphere to $SU_q(2)$ implies, 
 using the special properties of quantum groups at roots 
 of unity, that KK reduction on the quantum sphere 
 leads to a cutoff in the spins of KK states. 
 Using the match of KK states with {\it generating } chiral primaries
  \refs{\db}, the cutoff on such chiral primaries 
 in the CFT was analyzed and shown to be in agreement 
 with the above $SU_q(2) \times SU_q(1,1)$ spacetime. 
 Another way to arrive at the { \it same}   value of $q$ 
 was dynamical as in  similar  work in earlier matrix models \jv.   
 An $SL_q(2)$ subalgebra was identified
 in the algebra generated by the chiral primaries 
 and their conjugates. This makes it clear that 
 the parameter $1/N$ governing the non-commutativity
 of space-time is identical to the parameter measuring the strength 
 of interactions, so that the emerging model 
 of non-commutative gravity here does not involve
 doing classical gravity on a non-commutative spacetime
 followed by quantum effects being turned as another
 parameter is being varied. 
 A review of ideas in the general subject of 
 non-commutative space-times in the gravitational context,  
 with extensive references is given in \madore.

 The techniques of collective field 
 theory  allow,  in general,  the construction of spacetime actions 
 starting from the algebra of $S_N$ invariant variables. 
 Since the short representations related to chiral primaries
 have been matched with gravity on $ADS_3 \times S^3$, 
 it is instructive to focus on the dynamics associated 
 with these representations and their CFT conjugates. 
 We describe a central role played by  a Lie 
 algebra of observables associated to the chiral primaries. 
 The simplest representations of this Lie algebra 
 involve the Fock space generated by a certain class
 of chiral primaries  ( those which are generators
 of the chiral ring ). One wants to  focus on these representations 
 first because they include states created by the graviton, whose
 propagation  and dynamics is
 of interest. We describe this dynamics in terms of two
 dimensional field theory. This represents the simplest
 representation of the algebra.
 Other reps. will be of relevance when we go beyond chiral 
 primaries and study more stringy states 
 \refs{ \bhdm, \ml, \sw }.

The paper is organized as follows.
In ch.2,  we summarize the field  content
and basic symmetries of the CFT, and then 
 describe the computation of some exact 
 correlation functions involving twisted sector 
 chiral primaries. 
In ch.3 concentrating on the untwisted sector
 we review the algebra of observables given by
the creation-annihilation operators and their commutators.
 We study a Fock space representation of this 
 algebra where a creation operator 
 is associated to each generating  chiral primary.
 We use a coherent state description for the Fock 
 space and derive, using the correlation functions 
 of the chiral primaries, a formula for a
 symplectic form ( equivalently the kinetic term 
 of a Lagrangian ) on the space of coherent states. 
 We also obtain a formula for the Hamiltonian 
 in this coherent state representation. 
 We outline how these exact formulae can be developed 
 in a $1 \over N$ expansion into a realization of 
 the algebra in terms of free oscillators. 
 In section 4, we describe the algebra when twisted sector  
 operators are taken into account. We describe 
 the leading terms in a systematic ${1 \over N}$
 free field representation of the algebra. 
  We study the form of the Hamiltonian, and observe 
 that free field realizations which differ in detailed 
 form, lead to the possibility of a quadratic Hamiltonian 
 and an interacting one. We write a formula for the interacting Hamiltonian 
 and observe similarities with other gravity-gauge theory 
 correspondences, which suggest that the dynamics of chiral primaries
 can be understood as a reduction of $ADS \times S$ backgrounds 
  to a $1+1$ dimesnional 
 system having some universal features. 
In section $5$ we turn to finite $N$ effects,
like the stringy exclusion principle, 
 where the free field realizations start 
 to break down. We outline a non-trivial property 
 of the finite $N$ Lie algebra which follows from 
 independently known facts about the cohomology 
 of the instanton moduli spaces. We return to the picture
 of a non-commutative spacetime emphasized in I. The 
 q-deformed symmetry $SU_q(2|1,1)$ of the q-spacetime 
 proposed in I, is seen to govern the structure 
 of the fusion rules implicit in the correlation 
functions of section 2. While I focused on the quantum 
spacetime interpretation of the single-particle states, 
 we begin a  study of multi-particle states in this franework. 
  Finally we study the properties of the algebra 
 and look for the value of $L_0 $ where 
 deviations from free field behaviour first show up. 

We conclude with a summary and 
outline some future directions.

\newsec{Orbifold Conformal Field Theory. Field Content and Symmetries}

In this section we will discuss the field content and the symmetries of
the SCFT on symmetric  product  
$S^N(X)$, where $N=Q_{1}Q_{5}$ and $X$ is either $T^{4}$ or $K3$ . It
is known that this SCFT  has (4,4) 
superconformal symmetry in both cases. 
We will work with $T^{4}$ for simplicity. Many of the results 
extend  simply to  $K3$. 
\par
The field content of the theory is: $4N$ 
real free bosons $X^{a \dot{a}}_{I}$ representing the 
coordinates of the 
torus and their superpartners $4N$ free fermions 
$\Psi^{\alpha\dot{a}}_{I}$, where $I=1,..,N$, 
$\alpha,\dot{\alpha} =\pm$ are the spinorial $S^{3}$ 
indices, and $a,\dot{a}=1,2$ are the spinorial indices on $T^{4}$.
 Using the relation between the Fermi fields: 
$\Psi^{\alpha\dot{a}\,\dagger} = \Psi_{\alpha\dot{a}}=
\epsilon_{\alpha\beta}\epsilon_{\dot{a}\dot{b}}
\Psi^{\beta\dot{b}}$, the field content of the 
theory is determined to be $4N$ real free bosons 
and  $2N$ Dirac free fermions, 
giving a central charge $c=6N$. The  left moving superconformal 
symmetry is generated by the following currents 
\eqn\energ 
{
 T(z) = -{1 \over 2}\sum_{I}\,\partial X_{I}^{a\dot{a}} \, 
\partial X_{I\,a\dot{a}}-{1 \over 2}\sum_{I} \,
\psi_{I}^{\alpha\dot{a}} \,\partial \psi_{I \,\alpha\dot{a}}, 
}
\eqn\susy {
G^{\alpha a}(z) = i\sum_{I}\, \psi^{\alpha \dot{a}}_{I} \, 
\partial X_{I \, \dot{a}}^{a} ,
} 
\eqn\rsymm {
J^{\alpha\beta}(z) = {1 \over 2}\, \sum_{I} \, 
\psi_{I}^{\alpha\dot{a}} \, \psi^{\beta}_{I \, \dot{a}},
}
where $\psi^{\alpha\dot{a}}$ is the left moving component of 
the coresponding fermion, and the $\epsilon_{\alpha\beta}$,
$\epsilon_{\dot{a}\dot{b}}$ are used to raise and lower the 
spinorial indices. The lowest modes of this currents 
$ \{ L_{0,\pm 1},G_{\pm  {1 \over 2}}^{\alpha a},J_{0}^{\alpha\beta} \}$
 will generate together to their right counterparts the $SU(2|1,1)_{L} 
\times SU(2|1,1)_{R}$ symmetry which is mapped in the $AdS/CFT$ 
correspondence to the superisometries of the $AdS_{3}\times S^{3}$. 
In addition, it is possible to construct other symmetries commuting 
with the previous set and related to global $T^{4}$ rotations, and 
they are given by the following currents:
\eqn\torsymm
 {
K^{\dot{a}\dot{b}}(z)= {1 \over 2}\,\sum_{I}\,(\psi_{I}^{\alpha\dot{a}}\,
\psi_{I \, \alpha} ^{\dot{b}}-X_{I}^{a\dot{a}} \, \partial 
X_{I \, a}^{\dot{b}}),    
}
and similar expressions for the right-movers.
This symmetry acts non-trivially on the space of chiral primaries. 
\par
Although the underlying CFT on $T^4$ 
 is free, non-trivial $S_N$ invariant chiral primary operators
 can be constructed in correspondence  with conjugacy 
 classes of $S_N$ \dfms\Ver\Frol. 
The basic twist operators for $n$ free bosons: $X_{I}, I=1..n$, 
 are  defined by the OPE : 
\eqn\opg
{
\partial X_{I}(z)\sigma_{(1..n)}(0)=z^{{1 \over n}-1} e^{-{{2 \pi i } 
\over {n}}I}\tau_{(1..n)}(0)+..
}
They impose the boundary conditions : 
\eqn\bcond {
X_{I}(z\,e^{2 \pi i}, \bar{z}\,e^{-2 \pi i})=X_{I+1}(z,\bar{z}), \; 
I=1..n-1, \;
X_{n}(z\,e^{2 \pi i}, \bar{z}\,e^{-2 \pi i})=X_{1}(z,\bar{z})
}
The general twist operator for a general 
conjugacy class of $S_N$ is obtained 
 by its decompostion into cycles.
We may define a twisted sector vacuum by 
$|(1..n)\rangle=\sigma_{(1..n)}(0)|0\rangle$. 
Over this vacuum we can 
write the following mode expansion: 
\eqn\osc 
{
\partial X_{I}(z)=- {i \over n}\sum_{m}\,\alpha_{m}\,
e^{-{{2 \pi i} \over {n}}Im}z^{-{m \over n}-1}
}
where $[\alpha_{m},\alpha_{n}]=m\delta_{m+n,0}$ and 
$\alpha_{m}|(1..n)\rangle=0$ for $m \geq 0$.
The expression above 
can be generalized to any other primary operator constructed from $X_{I}$, 
the only difference being that 
in the exponent of $z$, 1 is replaced by the corresponding conformal 
dimension. This shows that the $S_{N}$ invariant operators have 
nonsingular OPE with 
the twist operator. The dimension of the twist operator can be 
calculated by computing the energy momentum tensor in the state 
$|(1..n)\rangle$ and it is (for one boson, left-mover):
\eqn\dimens {
\Delta_{(1..n)}={1 \over 24}(n-{1 \over n})
}  
\par
For our system, we will bosonize first the fermions by 
introducing $\Phi_{I}^{1,2}(z,\bar{z})=\phi_{I}^{1,2}(z)+
\bar{\phi}_{I}^{1,2}(\bar{z})$, the final theory being one of $6N$ 
free bosons and we will construct the chiral primaries using 
these bosons : 
\eqn\bosg {
\psi_{I}^{+ a}(z)=e^{i\phi_{I}^{a}(z)}, \; \psi_{I}^{-a}(z)=
\epsilon_{ab} \, e^{-i\phi_{I}^{b}(z)}
}
for left-movers and:
\eqn\bosp {
\bar{\psi}_{I}^{+a}(\bar{z})=e^{i\bar{\phi}_{I}^{a}(\bar{z})}, 
\; \psi_{I}^{-a}(\bar{z})=\epsilon_{ab}
\,e^{-i\bar{\phi}_{I}^{b}(\bar{z})}
}
for right-movers. Using these expressions,the $SU(2)_{L}$ currents 
are given in the standard way:
\eqn\rcurrent {\eqalign
{ & J^{3}={i \over 2}\sum_{I} \, (\partial \phi_{I}^{1}(z)+\partial 
\phi_{I}^{2}(z)),\cr 
  & J^{+}=\sum_{I}\,e^{i(\phi_{I}^{1}+\phi_{I}^{2})(z)},\cr
  & J^{-}=\sum_{I}\,e^{-i(\phi_{I}^{1}+\phi_{I}^{2})(z)}.\cr
}}

\par

Let us construct the $Z_n$ twist 
operators which can be used to build 
$S_N$ invariant chiral primaries by averaging over $S_N$.
These twist operators play a distinguished role 
in the chiral ring in the sense that they 
 can be used to generate the rest by using the ring structure. 
In the correspondence with gravity in $ADS_3 \times S^3$ 
they are in one-one correspondence with single particle states.   
These vertex operators are written in terms of the $6$ free scalar
fields and the twist operators. 
Consider first the fields appearing in the untwisted sector,
 $n=1$:
\eqn\untwp{\eqalign
{ & O_{(1)}^{(0,0)}=1, \cr
  & O_{(1)}^{a}=\psi_{1}^{+ a}, \cr
  & O_{(1)}^{\bar a}=\bar \psi_{1}^{+ a}, \cr
  & O_{(1)}^{a,\bar b }=\psi_{1}^{+ a} \bar \psi_{1}^{+ b}, \cr
  & O_{(1)}^{a a' ,\bar b }=\psi_{1}^{+ a} \psi_{1}^{+ a'}
    \bar \psi_{1}^{+ b}, \cr
  & O_{(1)}^{a ,\bar b \bar b' }=\psi_{1}^{+ a} \bar \psi_{1}^{+ b}
    \bar \psi_{1}^{+ b'}, \cr
  & O_{(1)}^{(2,2)}=\psi_{1}^{+ 1} \psi_{1}^{+ 2} \bar \psi_{1}^{+ 1}
    \bar \psi_{1}^{+ 2}. \cr
}}
where $(1)$ represents the one cycle containing only $I=1$.
We will also construct vertex operators out of the simple twists which
will
correspond to the nontrivial chiral fields associated  with the twist. For 
that we consider the cycle $(1..n)$ and we define the following $S_{n}$
invariant $6$ dimensional vector observables (left and right) : 
\eqn\vect { \eqalign {
 & Y_{L}(z)={1 \over n}\sum_{I=1..n}~(X_{I~L}^{1}, X_{I~L}^{2},
X_{I~L}^{3},
                               X_{I~L}^{4},
\phi_{I}^{1},\phi_{I}^{2})(z),\cr  
 & Y_{R}(\bar z)={1 \over n}\sum_{I=1..n}~(X_{I~R}^{1}, X_{I~R}^{2},
 X_{I~R}^{3},X_{I~R}^{4},\bar \phi_{I}^{1},\bar \phi_{I}^{2})(\bar z),\cr 
}}

Let us consider now the following field which consists 
of the twist operator for all 6 fields and having momenta along 
the 2 extra $\Phi$ dimensions:   
\eqn\vert {
O_{(1..n)}^{(0,0)}(z,\bar{z})=e^{i\sum_{I}\,({{n-1 } \over {2n}}
\phi_{I}^{1}+{{n-1 } \over {2n}}\phi_{I}^{2}(z)+{{n-1 } \over {2n}}
\bar{\phi}_{I}^{1}(\bar{z})+{{n-1 } \over {2n}}
\bar{\phi}_{I}^{2}(\bar{z}))}\,\sigma_{(1..n)}(\Phi,X)(z,\bar{z})
}
The dimensions of this field is $({{n-1 } \over {2}},{{n-1 } \over {2}})$ 
and its charges can be shown to be equal to the dimensions.
In comparison with the formulae in I, we have extracted 
and made explicit the $U(1)$ charges by exhibiting 
 exponentials of the bosons $\Phi$, leaving a twist
 operator for both the $\Phi$ and $X$. 
This field will be used to construct the $S_{N}$ invariant chiral 
primary $O_{n}^{(0,0)}(z,\bar{z})$ \malstro. A more precise way to write
which we will further generalize to the other cases is to write:
\eqn\vertg{
O_{(1..n)}^{(0,0)}(z,\bar{z})=e^{i (k_{L}Y_{L}(z)+k_{R}Y_{R}(\bar z) ) } 
             \sigma_{(1..n)}(\Phi,X)(z,\bar{z})
}
where $k_{L}=(0,0,0,0,{n-1 \over 2},{n-1 \over 2})$, 
$k_{R}=(0,0,0,0,{n-1 \over 2},{n-1 \over 2})$ represents the left 
and right momenta in $6$ dimensions. The dimensions of a field  of type
\vertg\ is given by the following general formula:
\eqn\dimg{\eqalign{
& \Delta_{O}=6{1 \over 24}(n-{1 \over n})+{1 \over  2 n} k_{L}^2, \cr 
& \bar \Delta_{O}=6{1 \over 24}(n-{1 \over n})+{1 \over  2 n} k_{R}^2, \cr 
}}
and for our $k$'s we obtain the dimensions above. For the charge we can
read it from the momenta on only the two extra dimensions, the twist being
uncharged. In order to construct all the other chiral fields we will
combine
the construction in untwisted sector with the twist. We will focus then on
the following construction:
\eqn\basc{
O_{(1..n)}^{A}(z,\bar z) \leftarrow 
O_{(1)}^{A}(z,\bar z)~O_{(1..n)}^{(0,0)}(z,\bar z)
}
where $A$ index takes care of the spinorial indices which already fully 
appear at untwisted level. We give for these operators a construction 
similar to \vertg\  where what makes the 
 distinction between them is the value 
for the momenta. This construction will be further justified by 
looking to the OPE of chiral fields.   

\par

We will focus further on scalar chiral primaries, namely those fields
coresponding to $O_{n}^{(0,0)}(z,\bar{z})$, 
$O_{n}^{(1,1)}(z,\bar{z})$, $O_{n}^{(2,2)}(z,\bar{z})$ only. 
We can
 characterize the fields as being a twist operator in 6 bosonic 
dimensions and having definite momenta on the bosonic fields coming
 from the bosonization $\Phi^{1,2}$ (the dimension and the charge for this
operators are equal):

\par 
$~O_{(1..n)}^{(0,0)}(z,\bar{z})$ coresponds to momenta
\eqn\momi{ 
k_{L}=(0,0,0,0,{{n-1 } \over {2}},{{n-1 } \over {2}}) ~~
k_{R}=(0,0,0,0,{{n-1 } \over {2}},{{n-1 } \over {2}}) } 
and has dimension $({n-1 \over 2},{n-1 \over 2})$,

\par

$~O_{(1..n)}^{(1,\bar{1})}(z,\bar{z})$ coresponds, for example, 
 to momenta 
\eqn\momii{ 
k_{L}=(0,0,0,0,{{n+1 } \over {2}},{{n-1 } 
\over {2}})~  k_{R}=(0,0,0,0,{{n+1 } \over {2}},{{n-1 } 
\over {2}})} 
$3$ other   combinations of left-right momenta 
in the case of $T^{4}$ are possible. The  dimension is 
$({n \over 2},{n \over 2})$. 
\par
$~O_{n}^{(12,\bar{1}\bar{2})}(z,\bar{z})$ coresponds to
momenta
\eqn\momiii{ 
 k_{L}=(0,0,0,0,{{n+1 } \over {2}},{{n+1 }
 \over {2}}) ~~  
  k_{R}=(0,0,0,0,{{n+1 } \over {2}},{{n+1 } \over {2}}) } 
and has dimension $({n+1 \over 2},{n+1 \over 2})$.

\par

We will define below the precise relation between the $O_{n}$ and 
the corresponding $O_{(1..n)}$
in a general context. For this let us consider the following basis of 
forms leaving on the target space $X$ 
and spanning its $H^{(1,1)}(X)$; we will denote them as
 $\omega^{r}_{a\bar{a}}$ where r counts the forms 
(for example, $r=1..4$ for $T^{4}$, and $r=1..20$ for $K3$)) 
and $a,\bar{a}=1,2$ and they are $X$ indices.
Using this forms and summing over all permutations, it is 
possible to describe the scalar chiral primaries up to
a normalization constant which will be determined in the next section. 
We write here the full expression for this operators: 
\eqn\twist{\eqalign
{ & O_{n}^{(0,0)}(z,\bar{z})={{1 } \over 
{(N!(N-n)!n)^{1 \over 2}}}\sum_{h\in S_{N}}
O_{h(1..n)h^{-1}}(z,\bar{z}),\cr
  & O_{n}^{r}(z,\bar{z})={{1 } \over {(N!(N-n)!n)^{1 \over 2}}}
\sum_{h\in S_{N}}O_{h(1..n)h^{-1}}^{a,\bar{a}}
\omega^{r}_{a\bar{a}}(z,\bar{z}),\cr
  & O_{n}^{(2,2)}(z,\bar{z})={1 \over 4}\,{{1 } 
\over {(N!(N-n)!n)^{1 \over 2}}}\sum_{h\in S_{N}}
O_{h(1..n)h^{-1}}^{ab,\bar{a}\bar{b}}\epsilon_{ab}
\epsilon_{\bar{a}\bar{b}}(z,\bar{z}),\cr
}}

\subsec{Correlation Functions}
 
We will compute in this section the OPE of two chiral 
primary operators focusing only on the constant appearing 
in front of the  
resulting single-particle chiral primary operator. 
From this we are able 
to write their three-point correlation functions. 
  Using the definition of the chiral 
primaries, we see that there are some restrictions coming from 
charge conservation, from  the fact that except the twist operator,
 these operators are fermionic in nature and  from $S_{N}$ 
multiplication law. It is straightforward to observe that we will be
 interested from the permutation point of view in two kind of processes: 
\par
1) the joining of two cycles overlapping on only one of their 
components giving a longer permutation, 
 which gives correlation functions between 
fields which have $(p,q)$ form indices as follows : 
i)$(0,0)+(0,0)\,\rightarrow \,(0,0)$,
ii)$(0,0)+r \rightarrow r$,
iii) $(0,0)+(2,2) \rightarrow (2,2)$,
iv)$s+r\,\rightarrow (2,2)$,
\par
2) the joining of two cycles overlapping on two components 
giving a longer permutation, which gives correlation functions 
of the form:
$(0,0)+(0,0)\,\rightarrow (2,2)$.

We will first study the processes listed above on the 
components of the chiral primaries and then we will sum over 
all permutation 
to recover the $S_{N}$ invariance. Using the notation 
of \Frol\ we will denote the left-moving twist operator 
having momenta as $O_{(1..n)}(k)$ where
$k$ is the momenta and we assume that 
they are normalized in the sense that their two-point 
takes the following form:
\eqn\verte {
\langle ~ O_h(k_{1})^{\dagger}(\infty)
O_g (k_{2})(0)\rangle = \delta_{h,g^{-1}}
\delta_{k_{1},k_{2}}
}

Using the same method, 
we will derive in appendix the OPE of a twist $(1..n)$ 
having momenta 
$k_{n}$ 
and the twist $(n\,n+1)$ having 
momenta $k_{2}$ involving only the twist $(1..n+1)$ 
(corresponding to the first kind of process) the result being:
\eqn\oper{\eqalign
{& O_{(n\,n+1)}(k_{2})(u)\,O_{(1..n)}
(k_{n})(0) = {{C(2,n|n+1;k_{2},k_{n}) } \over {z^{\Delta_{n+1}-\Delta_{2}-
\Delta_{n}}}}(O_{(1..n+1)}(k_{2}+k_{n})(0)+\cr
 & ~ ~ ~ ~ ~+O_{(1..n-1\,n+1\,n)}
(k_{2}+k_{n})(0)), \cr
}}
Since  we are dealing with operators which 
are chiral primaries one has the exponent of $z$ being $0$, 
and the $C(2,n|n+1;k_{2},k_{n})$ 
are determined in the appendix. 
We will write here the expression for $C(2,n|n+1;k_{2},k_{n})$ 
and we will see what they are in our processes 
(only left movers):
\eqn\opr{\eqalign
{ & O_{(n\,n+1)}(u)\,O_{(1..n)}(0) =  
{1 \over 2}\,({{n+1 } \over {n}})^{{1 \over 2}} 
( O_{(1..n+1)}(0) + O_{(1..n+1\, n)}(0)),\cr
  & O_{(n\,n+1)}(u)\,O^{1}_{(1..n)}(0)  = 
{1 \over 2}\,(O^{1}_{(1..n+1)}(0)+O^{1}_{(1..n+1\,n)}(0)),\cr
  & O_{(n\,n+1)}(u)\,O^{12}_{(1..n)}(0)  = 
{1 \over 2}\,({{n } \over {n+1}})^{1 \over 2}
(O^{12}_{(1..n+1)}(0)+O^{12}_{(1..n+1\,n)}(0)).\cr
}}
The method also gives us OPE for all the other cases 
involved in the list above case 
1) for a 2 twist and an $n$ twist overlapping over only 
one component. Knowing that the total twist is the 
product of the left and right moving twist and excluding 
out of the OPE those cases when the left and right twists 
do not coincide we see that the result of the OPE is 
essentially the square of what we have in the previous 
equations. We can also observe that an extrapolation 
to the $n=1$ case, when the twist is nothing than the 
identity operator results in the following OPE:
\eqn\opera{\eqalign
{ & O_{(1\,2)}(u)\,\psi^{+ a}(0)  = 
O^{a}_{(12)}(0),\cr
  & O_{(1\,2)}(u)\,\psi^{+1}\,
\psi^{+2}_{(1..n)}(0) = {1  \over \sqrt{2}} \,O^{12}_{(12)}(0).\cr
}}
\par
The equations above show that it is enough to figure out 
the OPE for the twist fields and then generalize to the 
other chiral primary fields.
We will use from this point the CFT rules to compute the 
OPE for any twist operators overlapping on only one 
component of the permutations. 
What we will do is to compute the OPE for 3 twist of 
type $(1..n-1)$, $(n-1\,n)$ and $(n..n+k-1)$ at different 
locations in the complex plane,
and by making different limits and an induction process we 
obtain the following results including now the right moving twist:
\eqn\operr {
O_{(n..n+k-1)}(u,\bar{u})\,O_{(1..n)}(0)  = 
{{n+k-1 } \over {2\,n\,k}}\,(O_{(1..n+k-1)}(0)+..),
}
where the dots are for the field corresponding to the 
other permutation obtained by multiplying the two 
permutations.
Using the equations above we derive the OPE for the 
processes 1) and we list them here:
\eqn\opea{\eqalign
{& O_{(n..n+k-1)}(u,\bar{u})\,O^{r}_{(1..n)}(0)= 
{{1 } \over {2\,k}}\,(O^{r}_{(1..n+k-1)}(0)+..),\cr
 & O_{(n..n+k-1)}(u,\bar{u})\,O^{12,\bar{1}
\bar{2}}_{(1..n)}(0)  = {{n } \over {2\,k\,(n+k-1)}}\,
(O^{12,\bar{1}\bar{2}}_{(1..n+k-1)}(0)+..),\cr
 & O^{r}_{(n..n+k-1)}(u,\bar{u})\,O^{s}_{(1..n)}(0)  
= {{-2 } \over {n+k-1}}\,\omega^{r}\,*\omega^{s}
(O^{12,\bar{1}\bar{2}}_{(1..n+k-1)}(0)+..).\cr
}}
Note that 
 if we extrapolate the above expressions for the 
case $n=1$ we have:
\eqn\justf{\eqalign{
& O_{(1..k)}(u,\bar{u})\,O^{r}_{(1)}(0)= 
{{1 } \over {2\,k}}\,(O^{r}_{(1..k)}(0)+..),\cr
 & O_{(1..k)}(u,\bar{u})\,O^{12,\bar{1}
\bar{2}}_{(1)}(0)  = {1 \over {2~k^{2}}}\,
(O^{12,\bar{1}\bar{2}}_{(1..k)}(0)+..).\cr
}} 
where for the definition of untwisted operators we use \untwp.

\par
In addition there is the 2) process which we 
will focus now on. Using the above rules there 
is a single OPE which has to be figured out and 
this is the one involving only two twists for the 
same 2 cycle (overlapping of 2), and then the result 
can be extended to the other permutations using the 
known OPE from the previous list and an induction process. 
Consider  the OPE of 3 operators: the twists and 
the one formed from all 4 adjoints of fermions and no twist. 
In the limit when we put together the twist and the fermions 
this process is nothing else than the one used for normalization
 of the twist of length $2$; in the limit when one puts together the twists 
and then the fermions and ask the question on what kind 
of operator would give this result one is led to consider
 the following OPE:
\eqn\oped {
O_{(12)}(u,\bar{u})\,O_{(12)}(0)  = 
\psi^{1}\psi^{2}\bar{\psi}^{\bar{1}}\bar{\psi}^{\bar{2}}(0) 
}
The final needed OPE is then:
\eqn\opel {
O_{(1..n)}(u,\bar{u})O_{(n\,n-1..n+k-2)}(0)={{1 } 
\over {n\,k\,(n+k-3)}}
(O^{12,\bar{1}\bar{2}}_{(1..\hat{n}..n+k-2)}(0)+..),  
}
where by $\hat{n}$ we mean that $n$ is 
missing in this permutation.
We have at this moment the 2 and 3-point 
functions for all fields which we will use 
in constructing the chiral fields.

\subsec{ Correlation functions of Chiral primaries. }

In this section we will put back the sums over permutations 
with appropriate normalization factors, which together with 
the 3 point functions deduced in the previous section will
 allow us to write the full 3-point functions for the chiral 
primaries.
For  $O_{n}^{(0,0)}$ we have : 
\eqn\deff {O_{n}^{(0,0)}(z,\bar{z})=const.\,
\sum_{h\in S_{N}} O_{h(1..n)h^{-1}}(z,\bar{z}).}
The two point function for individual twists is
given \verte\ with a constant to be determined. 
\eqn\norm {\langle \,O_{n}^{(0,0)\,\dagger}(\infty)\,
O_{n}^{(0,0)}(0)\rangle=const.^{2}\sum_{h_{1,2}\in S_{N}}\,
\langle\,O_{h_{1}(1..n)h_{1}^{-1}}^{(0,0)\,\dagger}(\infty)\,
O_{h_{2}(1..n)h_{2}^{-1}}^{(0,0)}(0)\rangle\, .}Because the 
2-point function for twists is normalized to $1$, the sums will 
be rearranged in other two sums: a sum over all permutations and
 a sum on only those permutation which leave a cycle invariant. 
The sum   over all permutation will give a factor of $N!$  whereas
 the second sum will give a factor of $(N-n)!\, n$ leading to the
 expressions already listed in equation \twist.
\par
For the three-point functions we will use the normalized chiral 
primaries operators and we will also show how one can determine 
them for only three $O^{(0,0)}$ and then list the full results
 for all the other cases. The conservation law for the R-symmetry 
suggest that the only possibility is having only the process listed
 in 1) namely a cycle having length $n+k-1$, one having length
 $n$ and one having $k$ and the individual permutations have to
 overlapp on one entry:
\eqn\thp {\eqalign
{ & \langle\,O_{n+k-1}^{(0,0)\,\dagger}(\infty)\,
O_{n}^{(0,0)}(1)\,O_{k}^{(0,0)}(0)\rangle= \cr
  & \; \;  const.\,\sum_{h_{1,2,3} \in S_{N}}\langle
O_{h_{1}(1..n+k-1)^{-1}h_{1}^{-1}}^{\dagger}(\infty)\, 
O_{h_{2}(n..n+k-1)h_{2}^{-1}}(1)\,O_{h_{3}(1..n)
h_{3}^{-1}}(0)\rangle,\cr
}}
where the $const.$ comes from the normalization of each chiral 
primary field.The individual terms which appear in the sum are 
nonzero only if:
\par
$h_{1}(1..n+k-1)^{-1}h_{1}^{-1}h_{2} (n..n+k-1)h_{2}^{-1}
h_{3}(1..n)h_{3}^{-1}=1$  
\par
and in this case they are all equal to the expressions derived 
in the previous section. By rearranging the previous permutation 
equation it is possible to compute the sums, namely:
\par
$(1..n+k-1)^{-1}h_{1}^{-1}h_{2} ((n..n+k-1)h_{2}^{-1}h_{3}(1..n)
h_{3}^{-1}h_{2} ) h_{2}^{-1}h_{1} =1$. 
\par
 Then the result will be a $N!$ summing over $h_{1}$, a factor 
of $(N-k)!/(N-k-n+1)!$ coming from the possibilities of 
constructing a long permutation out of 2 small permutation, 
a factor of $(N-n)!\,n$   coming from the sum over $h_{2}^{-1}h_{3}$ 
which leave the $n$ cycle invariant and a factor of $(N-n-k+1)!\,(n+k-1)$ 
coming from $h_{1}^{-1}h_{2}$ which leave the full $n+k-1$ cycle
invariant. 
In addition there is a factor of $2$ coming from the two possibility 
of multiplying permutations in individual three-point functions. 
We will list below the final result gathering all factors and also 
the results for all nonzero nontrivial three-point functions of 
three chiral primaries:
\eqn\thpf{\eqalign
{ & \langle\,O_{n+k-1}^{(0,0)\,\dagger}(\infty)\,
O_{k}^{(0,0)}(1)\,O_{n}^{(0,0)}(0)\rangle =  
({{(N-n)!\,(N-k)!\, (n+k-1)^3 } \over 
{(N-(n+k-1))!\,N!\,n\,k}})^{{1 \over 2}},\cr
  & \langle\,O_{n+k-1}^{r\,\dagger}(\infty)\,
O_{k}^{(0,0)}(1)\,O_{n}^{s}(0)\rangle= 
({{(N-n)!\,(N-k)!\, n\,(n+k-1) } \over {(N-(n+k-1))!\,
N!\,k}})^{{1 \over 2}}\delta^{r\,s}\cr
  & \langle\,O_{n+k-1}^{(2,2)\,\dagger}(\infty)\,
O_{k}^{(0,0)}(1)\,O_{n}^{(2,2)}(0)\rangle=
({{(N-n)!\,(N-k)!\, n^{3} } \over {(N-(n+k-1))!\,N!\,k\,
(n+k-1)}})^{{1 \over 2}}\cr
  & \langle\,O_{n+k-1}^{(2,2)\,\dagger}(\infty)\,
O_{k}^{r}(1)\,O_{n}^{s}(0)\rangle=
-({{(N-n)!\,(N-k)!\, n\,k } \over {(N-(n+k-1))!\,N!\,
(n+k-1)}})^{{1 \over 2}}\omega^{r}\,*\omega^{s}\cr
  & \langle\,O_{n+k-3}^{(2,2)\,\dagger}(\infty)\,
O_{k}^{(0,0)}(1)\,O_{n}^{(0,0)}(0)\rangle=\cr 
  & \; \; \; \; \; =\,2\,({{(N-n)!\,(N-k)!\,(N-(n+k)+3) } 
\over {(N-(n+k-1))!\,N!\,(N-(n+k)+2)\,n\,k\,(n+k-3)}})^{{1 \over 2}}\cr
}} 
Fixing the charges and taking 
$N \rightarrow \infty$:
\eqn\opear{\eqalign
{ &  \langle\,O_{n+k-1}^{(0,0)\,\dagger}(\infty)\,
O_{k}^{(0,0)}(1)\,O_{n}^{(0,0)}(0)\rangle=
({1 \over N})^{{1 \over 2}}({{(n+k-1)^3 } \over 
{n\,k}})^{{1 \over 2}},\cr
  & \langle\,O_{n+k-1}^{r\,\dagger}(\infty)\,
O_{k}^{(0,0)}(1)\,O_{n}^{s}(0)\rangle=
({1 \over N})^{{1 \over 2}}
({{ n\,(n+k-1) } \over {k}})^{{1 \over 2}}\delta^{r\,s} \cr
  & \langle\,O_{n+k-1}^{(2,2)\,\dagger}
(\infty)\,O_{k}^{(0,0)}(1)\,O_{n}^{(2,2)}(0)\rangle
=({1 \over N})^{{1 \over 2}}({{n^{3} } \over
 {k\,(n+k-1)}})^{{1 \over 2}}\cr
  & \langle\,O_{n+k-1}^{(2,2)\,\dagger}(\infty)\,
O_{k}^{r}(1)\,O_{n}^{s}(0)\rangle=
-({1 \over N})^{{1 \over 2}}({{n\,k } \over
 {(n+k-1)}})^{{1 \over 2}}\omega^{r}\,*\omega^{s}\cr
  & \langle\,O_{n+k-3}^{(2,2)\,\dagger}(\infty)\,
O_{k}^{(0,0)}(1)\,O_{n}^{(0,0)}(0)\rangle=2\,
({1 \over N})^{{1 \over 2}}({{1 } \over 
{n\,k\,(n+k-3)}})^{{1 \over 2}}\cr
}}
After the following rescalings 
\eqn\redef{\eqalign{ 
  & O_{n}^{(0,0)} \rightarrow {1  \over n}O_{n}^{(0,0)},\cr
  & O_{n}^{(0,0)\,\dagger} \rightarrow \, n\,
O_{n}^{(0,0)\,\dagger},\cr
  & O_{n}^{r} \rightarrow O_{n}^{r},\cr
  & O_{n}^{r\,\dagger} \rightarrow O_{n}^{r\,\dagger},\cr
  & O_{n}^{(2,2)} \rightarrow n\,O_{n}^{(2,2)},\cr
  & O_{n}^{(0,0)\,\dagger} \rightarrow {1 \over n}\,
O_{n}^{(0,0)\,\dagger},\cr
}}
which preserve the two-point functions, we have
three-point functions :
\eqn\opra{\eqalign
{ & \langle\,O_{n+k-1}^{(0,0)\,\dagger}(\infty)\,
O_{k}^{(0,0)}(1)\,O_{n}^{(0,0)}(0)\rangle=({{1 } 
\over {N}})^{{1 } \over {2}}((n+k-1)\,n\,k)^{{1 } \over {2}},\cr
  & \langle\,O_{n+k-1}^{r\,\dagger}(\infty)\,
O_{k}^{(0,0)}(1)\,O_{n}^{s}(0)\rangle=
({1  \over N})^{{1  \over 2}}((n+k-1)\,n\,k)^{{1 \over 2}}
\delta^{r\,s},\cr
  & \langle\,O_{n+k-1}^{(2,2)\,\dagger}(\infty)\,
O_{k}^{(0,0)}(1)\,O_{n}^{(2,2)}(0)\rangle=
({1  \over N})^{{1 } \over {2}}((n+k-1)\,n\,k)^{{1 } \over {2}},\cr
  & \langle\,O_{n+k-1}^{(2,2)\,\dagger}(\infty)\,
O_{k}^{r}(1)\,O_{n}^{s}(0)\rangle=-({1  \over N})^{{1 } 
\over {2}}((n+k-1)\,n\,k)^{{1 } \over {2}}\omega^{r}\,*\omega^{s},\cr
  &
\langle\,O_{n+k-3}^{(2,2)\,\dagger}(\infty)\,O_{k}^{(0,0)}(1)\,O_{n}^{(0,0)}(0)\rangle=2\,({{1
} \over {N}})^{{1 } 
\over {2}}((n+k-3)\,n\,k)^{{1 } \over {2}}\cr
}}
We notice  a certain degree of universality: 
all the three point
functions exibited above for the case of AdS3 have the same form factor
$\sqrt {nk(n+k)}$
as the three point functions  of chiral primaries in other AdS
examples for example 
\refs{ \mathur, \seiberg ,\zucchini }. We will have further
comments on this universal behaviour and its meaning in subsequent
discussion. 
In the future sections, we will denote 
as $A_{-n}^{(p,q)}$ the modes 
$ \sum O_{-\Delta(n,p,q)} \bar O_{-\Delta(n,p,q)} $,  
of $O_{n}^{p,q}(z,\bar z)$  
and by $A_{n}^{p,q}$ the conjugates.

\newsec{ Algebra of Observables 
and the Lagrangian - Untwisted sector }

At the heart of the disscusion begun in I 
is consideration of the commutator 
algebra generated by the exact single particle creation and 
annihilation operators $A_n^{p,q}$ and
$( A_n^{p,q})^+$ ( with the cuoffs on $n$ as described 
 in I ). 
These operators represent exact eigenstates of the Hamiltonian
and the full algebra is then by definition a spectrum generating one.
 We should 
 emphasize the analogy of this with previous large algebras appearing
 in related contexts: the W-algebra of the matrix
 model \refs{ \aj,\mpy}  the BPS algebras  of
\moorebps,  and extended algebras mentioned in  \vafa. 
An  interesting example of a spectrum generating algebra
in  semiclassical AdS gravity is discussed
 in \ggm.
  
In this section we concentrate
on the simplest, zero twist sector of the theory. It represents
a reduction of  CFT to a fermionic quantum mechanical system. We describe
for this the algebra of chiral primaries and the manner in which 
this algebra defines the interacting Lagrangian.
The collective \js\ interacting theory emerges as a representation of the 
algebra in terms of invariants. The relevance of the algebra on the 
Poisson (phase space)  structure of the theory will 
also be elaborated. 

  Let us begin from  the operators representing the
 chiral primaries $ A_{-1}^{(p,q)} $
 and their conjugates $ A_{1}^{p,q}$.  
 The procedure to construct the associated super-Lie 
 algebra is to consider 
 the ( graded ) commutators of the 
 conjugates with the chiral primaries. 
 The terms appearing on the RHS of the commutators
 are then commuted with the chiral primaries and their
 conjugates, and with each other until closure is achieved.
 This clearly generates a finite dimensional
 super-Lie algebra of which the creation-anihilation
operators and the hamiltonian are members. 
 A set of mutually commuting hermitian operators
 in this algebra are  
\eqn\cart{\eqalign{ 
& \psi^{\mu}_{I} ( \psi^{\mu}_{I})^{\dagger} \cr 
& ( \psi^{\mu}_{I} \psi^{\nu}_{I} ) ( \psi^{\mu}_{I} \psi^{\nu}_{I} )^
{\dagger } \cr 
&  ( \psi_{I}^{\mu} \psi^{\nu}_{I} \psi^{\sigma}_{I} )    
    ( \psi_{I}^{\mu} \psi^{\nu}_{I} \psi^{\sigma}_{I} )^{\dagger } \cr }}
 The indices $\mu$, $\nu$, $\sigma$ above run from $1$ to $4$ 
 and we have $4$ operators in the first line, 
 $6$ in the second line, and $ 4$ in the third line. 
 We should add to this an operator 
 $E$ which commutes with everything, 
and which appears,  for example,  in the commutator 
 of a an operator $A_{-1}^{(1,0)} $ and its conjugate.  
 Note that the procedure of successive 
 commutations of the $A_{-1}$ operators do not 
 generate terms involving a product of four 
 fermions with their conjugates. 

The other generators of the Lie algebra 
include of course the chiral primaries and conjugates, 
 along with others generated by the commutations. 
We also have the following operators and their conjugates. 
\eqn\morops{\eqalign{ 
&  \sum_{I} \psi_{I}^{\mu_1} ( \psi_{I}^{\nu_1})^{\dagger} \cr 
&   \sum_{I} \psi_{I}^{\mu_1} \psi_{I}^{\mu_2} ( \psi_{I}^{\nu_1})^{\dagger} 
\cr 
&   \sum_{I} \psi_{I}^{\mu_1} \psi_{I}^{\mu_2} 
( \psi_{I}^{\nu_1} \psi_{I}^{\nu_2} )^{\dagger} \cr 
&    \sum_{I} \psi_{I}^{\mu_1} 
\psi_{I}^{\mu_2} \psi_{I}^{\mu_3} ( \psi_{I}^{\nu_1} )^{\dagger} \cr 
&  \sum_{I} \psi_{I}^{\mu_1} \psi_{I}^{\mu_2} \psi_{I}^{\mu_3} 
          ( \psi_{I}^{\nu_1} \psi_{I}^{\nu_2} )^{\dagger} \cr 
&  \sum_{I} \psi_{I}^{\mu_1} \psi_{I}^{\mu_2} \psi_{I}^{\mu_3}
           ( \psi_{I}^{\nu_1} \psi_{I}^{\nu_2}  \psi_{I}^{\nu_3} )^{\dagger}
 \cr
& \sum_{I}  \psi_{I}^{\mu_1} \psi_{I}^{\mu_2} \psi_{I}^{\mu_3}  
        \psi_{I}^{\mu_4} ( \psi_{I}^{\nu_1} )^{\dagger } \cr 
& \sum_{I}  \psi_{I}^{\mu_1} \psi_{I}^{\mu_2} \psi_{I}^{\mu_3}  
\psi_{I}^{\mu_4} 
           ( \psi_{I}^{\nu_1}\psi_{I}^{\nu_2} )^{\dagger}\cr
&\sum_{I}  \psi_{I}^{\mu_1} \psi_{I}^{\mu_2} \psi_{I}^{\mu_3}  
\psi_{I}^{\mu_4}( \psi_{I}^{\nu_1}\psi_{I}^{\nu_2}\psi_{I}^{\nu_3} )^{\dagger}
 \cr  }}

Some of these operators have already been included in the 
 description of the Cartan above. The operator of the  form
  $ ( \psi )^4 ( \psi^{\dagger} )^4 $ does not appear in the 
 the commutators. Let us call this 
  Lie super-algebra $g_{inv}$, and the corresponding supergroup
 $G_{inv}$. 
   If we nevertheless include it 
 we get a superlagebra which has rank 
 $16$ and is closely related  to the clifford algebra 
 which is in turn related to $SU(16)$ \foot{ We thank 
 J. Gervais for this remark. }. 
 This suggests that the Lie super-algebra is actually 
 $U(8|8)$.

\subsec{Coherent State Representation}
We derive in this section the action governing the dynamics of the chiral
primaries for the untwisted sector and then comment on how it is
extended to the full set of chiral primaries and beyond. 
Consider then the simplified 
model obtained by reducing 
the theory to a quantum mechanical version with the algebra
described above.
 We deal then with the following set of operators as annihilation operators:
\eqn\bop{ \psi^{\pm \,\mu}_{I};\; \; \mu=1..4, \; \; I=1..N,}
where $\pm$ is the $SU(2)$ index, $\mu$ is the $T^{4}$ index and we also have
the adjoints of these as creation operators. We also focus on the chiral
primaries only meaning that we consider 
only highest weight states under the 
action of the $SU(2)$, namely $(+)$, and for now on we will also drop this
index. The hamiltonian of the system (the one involving only highest weight
under $SU(2)$) is considered to be the hamiltonian of 
a free system of fermions:
\eqn\hamilt{
L_{0}+ {\bar L_0}  ={1 \over 2} \sum_{I \, \mu } 
\psi^{\mu \, \dagger}_{I} \, \psi^{\mu}_{I}
}

The chiral primary operators in the untwisted sector are built as:
\eqn\oper{\eqalign{ 
 & A_{-1}^{\mu }=
{1 \over \sqrt{N}} \sum_{I}  \psi^{\mu  \dagger}_{I}, \cr 
& A_{-1}^{\mu \nu }=
{1 \over \sqrt{N}} \sum_{I}  \psi^{\mu  
\dagger}_{I} \psi^{\nu  \dagger}_{I}, \cr
& A_{-1}^{\mu \nu \sigma  }=
{1 \over \sqrt{N}} \sum_{I}  \psi^{\mu \dagger}_{I}  
\psi^{\nu \dagger}_{I}  \psi^{\sigma  \dagger}_{I}, \cr
& A_{-1}^{\mu \nu \sigma \rho }={1 \over \sqrt{N}} 
\sum_{I}  \psi^{\mu  \dagger}_{I}
\psi^{\nu  \dagger}_{I} \psi^{\sigma 
\dagger}_{I} \psi^{\rho  \dagger}_{I}.\cr
}}  
The idea of the construction is to build 
them using only the highest weight operators under $SU(2)$ and $S_{N}$
 invariant combination of operators. The  dimension and 
the charge of the operators are equal to the number of 
fermion operators.
\par
Let us study now how we write an action involving only these special set of
operators. For this we can  use the technique  of \ber. It consists in introducing
a basis of coherent states using the $S_{N}$ invariant operators already 
derived and from the expression for the partition function we derive the 
hamiltonian and the symplectic form which governs the dynamics of the 
collective variables. The first step is to derive the measure which will be 
use together with the coherent basis $\mu (\xi_A , \bar \xi_A ) $ as: 
\eqn\defmeas{ \mu( \xi_B,  \xi_B^{*} )  = 
< 0 | e^{ \xi_B^{*} A_{1}^B }  e^{ \xi_B A_{-1 }^B } |0 > } 
Here we have taken the index $B$ to run over all the $15$ chiral primaries 
of the invariant sector. 
This can be computed using coherent state techniques to find : 
\eqn\ans{ \mu( \xi_A,  \xi_A^{*})= (Z( \xi , \xi^{*} ))^{N},}
where
\eqn\ansp { Z[ \xi, \xi^*] =   
 1 - {1 \over N } \tilde \xi^{\mu * } \tilde \xi^{\mu} + 
 { 1 \over 2 N} ( \tilde \xi^{\mu \nu } )^* ( \tilde \xi)^{\mu \nu }
 - { 1 \over 3! N} ( \tilde \xi^{ \mu \nu \rho} )^* 
    ( \tilde \xi^{ \mu \nu \rho} )       + { 1 \over 4! N}    
 ( \tilde \xi^{\mu \nu \rho \sigma}  )^* \tilde \xi^{\mu \nu \rho \sigma } } 
 is the answer for one spieces of fermions.
Here the following variables are used:
\eqn\deftild{\eqalign{
& \tilde \xi^{\mu}=\xi^{\mu},\cr
& \tilde \xi^{\mu \nu}=\xi^{\mu \nu}-{1 \over \sqrt{N}} \xi^{\mu} \xi^{\nu},\cr
& \tilde \xi^{\mu \nu \sigma}=\xi^{\mu \nu \sigma}+{3 \over \sqrt{N}}
       \xi^{\mu \nu} \xi^{\sigma}-{1 \over N} \xi^{\mu} \xi^{\nu} \xi^{\sigma},
\cr
& (\tilde \xi^{(2,2)})=\xi^{(2,2)}+{1 \over {3! \sqrt{N}}} 
      \epsilon_{\mu \nu \sigma \rho}
       \xi^{\mu \nu \sigma} \xi^{\rho}+ {1 \over {2! \sqrt{N}}}
        \epsilon_{\mu \nu \sigma \rho} \xi^{\mu \nu} \xi^{\sigma \rho}+...,\cr
}}
and the dots stand for terms of lower order in ${1\over N}$ terms.    
\par
For determining the representation of the hamiltonian in $\xi$, $ \xi^*$ we 
evaluate its matrix elements in the coherent basis. In our case it is 
suitable to evaluate the following:
\eqn\matel{ 
Z[\beta,\xi, \xi^*]=<0 | e^{\xi^*_B 
A_{1}^B } e^{ \beta ( L_0 + \bar L_0)  } 
e^{ A_{-1 }^B \xi_B } |0>  
 }
Using the expression for the hamiltonian \hamilt and coherent state tehniques we
compute the following expression:

\eqn\prop{\eqalign{ 
 & Z [ \beta,  \xi,   \xi^* ] =  \cr 
 & Z [ ( 1 + { 1\over 2} \beta ) \tilde \xi^{\mu},
    ( 1 + { 1\over 2} \beta )  \tilde \xi^{\mu *} ; 
        (1 + 1/2 \beta )^2 \tilde \xi^{\mu \nu },   
         (1 + 1/2 \beta )^2  \tilde \xi^{\mu \nu * }  ; \cr  
&          (1 + 1/2 \beta )^3 \tilde \xi^{\mu \nu \rho }; 
         ( 1+ 1/2 \beta )^3  \tilde \xi^{\mu \nu \rho * } ; 
         ( 1 + 1/2 \beta )^4 \tilde \xi^{\mu \nu \rho \alpha };
           ( 1+ 1/2 \beta )^4  \xi^{\mu \nu \rho \alpha * } ] \cr }}
We can write then the average for the hamiltonian as: 
\eqn\ave{\eqalign{
  H (  \xi ,  \xi^* ) & \equiv
 {  < \bar   \xi| ( L_0 + \bar L_0)    | \xi > 
        \over { \mu ( \tilde \xi,  \tilde \xi^* ) }} \cr 
    &= { 1 \over Z^N} \partial_{\beta} Z^{N} |_{\beta = 0  } \cr
    & = { N \over Z } { \partial Z \over \partial \beta }|_{\beta =0 } \cr  
     & = {  { N \over 2 } ( {-1 \over N} 
        \tilde \xi^{\mu * } \tilde \xi^{\mu}    + \cdots ) \over 
             ( 1 - {1\over N }
           \tilde \xi^{\mu * } \tilde \xi^{\mu} + \cdots  ) }    \cr }}
In the last line we wrote the first quadratic term, the other being also 
quadratic in $\tilde \xi$ but for the remaining indices.
Noting the expression for the tilde variables (eq.3.9) one has a sequence of 
cubic,quartic and higher terms which are explicitely determined by
the above representation .
\par 
Using the equations above we can now derive the 
lagrangian governing the dynamics 
of the collective fields we introduced:
\eqn\lagran{
L(\xi, \xi^*)=L_{\omega}(\xi, \xi^*)-H(\xi, \xi^*),
}
 $L_{\omega}$ is determined from the above measure \ans ,\ansp\ by 
\eqn\formlag{ 
 L_{\omega}( \xi, \xi^* )  = 
    \bigl[ 
 { \partial \xi^{\mu }   \over \partial t  } 
          { \partial \over \partial \xi^{ \mu }   } 
              + 
      { \partial \xi^{\mu \nu }   \over \partial t  } 
          { \partial \over \partial \xi^{ \mu \nu }   } 
              + 
        { \partial \xi^{\mu \nu \alpha }   \over \partial t  } 
          { \partial \over \partial \xi^{ \mu \nu \alpha }   }b
              + 
         { \partial \xi^{\mu \nu \alpha \beta  }   \over \partial t  } 
          { \partial \over \partial \xi^{ \mu \nu \alpha  \beta }   }
         \bigr] \log \mu ( \xi ,  \xi^*), } 
and it is linear in time derivatives. Its expression gives the Poisson 
structure in the phase space of $\xi$, $\xi^*$. For the other part of the 
lagrangian $H$, we use the expression we computed in \ave.

\par 
 This coherent state technique 
 can also be used to give a free field realization 
 of the algebra of observables in a $1/N$ expansion 
 which reproduces correlation functions involving 
 a small number of operators compared to $N$. 
 To convert to free fields we have to find  
  variables which  convert the Lagrangian 
  $L( \xi , \xi^*) $ into $ L( a, a^*) = 
  a { \partial  a^* \over \partial t }$. This can be done 
 in a systematic large $N$ expansion and reproduces 
 correlators involving a small number of insertions 
 compared to $N$. 
 When the number of insertions becomes comparable to $N$, 
 effects like the stringy exclusion principle become 
 important. The properties of the measure 
  $ \mu $ then show properties qualitatively 
 diferent from free fields. For example 
\eqn\dervan{ 
 \bigl( {  \partial \over  \partial \tilde \xi } 
     { \partial \over \partial \tilde \xi^* }\bigr) ^{N=1} 
   \mu ( \xi, \tilde \xi^* ) =  0   } 
 The coherent state technique offers a complementary insight 
 into the exclusion principle. It is related to the fact that  
 the symplectic manifold generated by the action of 
 elements of $G_{inv}$ on the Fock vacuum has a non-trivial 
 symplectic form which cannot   be globally brought 
 to the form $ \omega = \sum_i dp^i \wedge dq^i $. 
 This phenomenon was emphasized in simpler models of large $N$
 in \ber. 
 We expect along the lines of I, that a transformation valid
 for finite $N$ can be done using q-oscillators at roots 
 of unity.

\newsec{ Algebra of observables - Twisted sector } 
 
 The strategy for defining a Lie algebra 
 associated with the single particle 
 chiral primaries by taking successive 
 commutators clearly works when we include the twisted sectors  
 as well. Some elements were described in 
 I.  Here we will discuss further properties of the 
 algebra, and its representations in terms of free fields.
 As we go from the zero twist to the twisted sector the
 the creation-anihilation operators aquire an additional index $n$,
 which is reflected in the commutator algebra. The commutators 
 also become more complex, but are computable 
 using CFT techniques. The general form of the 
 terms are highly constrained by conservation of $L_0,J_0$, 
 and the $S_N$ symmetry. The $N$ dependences 
 can be obtained by $S_N$ combinatorics.  
  The precise coefficients 
 are related to computations of correlation functions 
 of the kind done in section 2.

 Consider the class of chiral primary operators 
 $ A_{-n}^{0,0}$ defined at the end 
of section 2,  denoted here  $A_{-n}$. 
By elementary commutator manipulations,  followed 
 by contour integrals, we derived in I an
 equation  which reduces in the large $N$ limit
to 
\eqn\algi{\eqalign{ 
  [ A_{-n} , A_{-n}^{\dagger}  ] 
  =  1 + { 1 \over N^{n \over 2}} 
  [ \sum_{ \sigma_n \in T_n }  
    O^{\sigma_n}O^{ \dagger \sigma_n }   + 
   {\bar O}^{\sigma_n}
 {\bar O}^{ \dagger \sigma_n }  ]  } } 
We are keeping only the leading 
 terms in the combinatoric factors, e.g $ { N\choose n}$
 has been approximated by $N^{n}$, to demonstrate
 the key features of the large $N$ counting. 
Each operator is normalized to $1$. 

We will now consider the 
form of commutators involving
different  chiral primaries $A_{-m_{1}}$ and $A_{-{m_2} }^{\dagger}$. 
For this we will need to discuss
commutators of the chiral building blocks 
\eqn\chircom{ 
 [ O^{\sigma_{m_1} }  , O^{\sigma_{m_2}\dagger}  ] 
 = C^{\sigma_{m_1} \sigma_{m_2} }_{\sigma_{3} } O_{\sigma_{3} } 
} 
 In the above equation 
 $\sigma_{m_i} \in T_{m_i}$ where $T_{m_i}$ denotes 
 the set of permutations in $S_N$ which have
 one cycle of length $m_i$ and remaining cycles of length  
 $1$. We are not being specific about the order 
 of magnitude of $C$ in the large $N$ expansion 
 in the above equation, but we will cure  that 
 in a moment. 
 We will assume that $m_2$ is larger than $m_1$. 
 We can further decompose 
 the sum over $\sigma_3$ by specifying the 
 number of non-trivial cycles ( of length greater
 than $1$ ) in the permutation $\sigma_3$. 
 When $\sigma_3$ belongs to a permutation with 
 $k$ non-trivial cycles of lengths $( n_1, n_2, \cdots n_k )$
 we say $ \sigma_3 \in T_{n_1, n_2, \cdots n_k }$, 
 and we write $\sigma_3 $ as $\sigma_3^{n_1} 
\sigma_3^{(n_2)} \cdots \sigma_3^{(n_k)}$. In the following equation 
 we have normalized the operators to have 2-point functions 
 equal to one. The $C$-factors are of order one. And the form 
 of the leading $N$ dependences have been 
 made explicit. 
\eqn\chircomi{\eqalign{  
& \sum_{\sigma_{m_1} \in T_{m_1}, \sigma_{m_2} \in T_{m_2} } 
[ O^{\sigma_{m_1}}  , O^{\sigma_{m_2}\dagger}  ] ~~= \cr 
& \sum_{\sigma_3 \in T_{n_1}} N^{-{ 1\over 2}} 
  ( 1 + { \cal O }( { 1 \over N} ) ) 
  \delta ( m_2 - m_1 + 1 , n_1) 
 C^{\sigma_{m_1} \sigma_{m_2} }_{\sigma_{n_1} } 
      O^{\dagger \sigma_{n_1}}\cr
&  +  \sum_{\sigma_3 \in T_{n_1, n_2 } }
       N^{-1}  
        ( 1 + { \cal O } ( { 1 \over N } ) )  
    \delta ( m_2 - m_1 + 2, n_1 + n_2 )
 C^{\sigma_{m_1} \sigma_{m_2} }_{\sigma_{3} } O^{\dagger \sigma_{n_1} } 
    O^{ \dagger \sigma_{n_2} }  \cr 
& + \cdots \cr 
& + \sum_{\sigma_3 \in T_{n_1, n_2, \cdots n_k } } N^{-k/2}
      ( 1 + { \cal O }( {1 \over N} ) ) 
      \delta ( m_2 - m_1 + k,  n_1 + n_2 + \cdots n_k ) \cr 
&       C^{\sigma_{m_1} \sigma_{m_2} }_{\sigma_{3} } 
     O^{ \dagger \sigma_{n_1}} 
     O^{\dagger\sigma_{n_2}}  \cdots O^{\dagger \sigma_{n_k} } \cr }}
Note that we could have used a basis  
 where the product operators are just products 
 of the generating $S_N$ invariant generating 
 chiral primaries. For example we could have written 
\eqn\altbad{ 
  \sum_{ \sigma_{1}  \in T_{n_1}, \sigma_2 \in T_{n_2}  } 
   O^{\sigma_{1}}  O^{\sigma_{2}} } 
 rather than the sums 
 associated with conjugacy classes as in \chircomi. 
 In the sum in \altbad\ the terms $\sigma_1$ and $\sigma_2$
 can involve elements which overlap while they do not 
 involve overlapping elements in \chircomi. 
 The advantage of writing the algebra in the form above is that 
 the leading coefficient can be read off directly from 
 the 3-point functions that have been computed in
section 2. 
 
 Another noteworthy point in \chircomi\ is that 
 after correct normalization the terms involving 
 permutations with higher numbers of non-trivial  
 cycles are sub-leading in the large $N$ expansion.  
 This means that by restricting attention 
 to leading powers of ${ 1 \over N} $ we can define 
 appropriate { \it contractions } of the Lie algebra of 
 interest which can be simpler than the exact Lie algebra 
 and may be investigated in connection with the qualitative 
 properties 
 such as  the 
 exclusion principle and the related properties 
of correlation functions
 which we discuss in  section 5.  
 
 While it is easy to prove that the full set of operators
 appearing in the commutator in \algi\ is 
 of the form given there, 
 we have not proved that the above form 
 of operators is the full set of terms that can appear in the 
 commutator in \chircomi. 
  We have some evidence that 
 the form of the algebra, restricting attention 
 to pure twists is of the above form. 
 For example we know that the RHS cannot contain 
 a descendant of a chiral primary, which would not satisfy 
 $ L_0 = J_0 $ as the LHS does. The $L_0$ and $J_0$ conservations 
 alone do allow other forms of operators which can be  
 ruled out by using $S_N$ selection rules. 
 In particular, it appears  that we do not need
 to include operators of the form $ ( O O^{\dagger} ) O^{\dagger} $.
 The generalization of the above ansatz to include the chiral factors
 which come up in the $A_{n}^{p,q}$ is easily 
 done at the cost of some extra notation. 

 We will be interested in a  class of 
  free field  realizations
 of the algebra obtained by considering the 
 full non-chiral operators. We note nevertheless 
 that similar free oscillator
 realizations can be considered 
 for the chiral half of the algebra 
 and that has some similarities with the large $N$ 
 counting relevant for the full operators. 
 We can realize the algebra in terms of free
 oscillators,
 with $[ \alpha_{-n}, \alpha_{-m} ] = \delta_{n,m} $.
 In the large $N$ expansion,  for example, by 
\eqn\frosc{\eqalign{& 
A_{-m} = \alpha_{-m} + { 1 \over \sqrt N } ( 1 + { \cal O }({ 1 \over N} ) )
                \sum_{m_1,m_2}  \alpha_{-m_1} \alpha_{-m_2} 
                       \delta ( m_1 + m_2 - 1, m  )   \cr 
               &  + \cdots + 
                { 1 \over N^{k/2}  } ( 1 + { \cal O }  ({ 1 \over N} ))
                  \sum_{m_1,m_2 \cdots m_k } \alpha_{-m_1} 
                      \alpha_{-m_2} \cdots \alpha_{-m_k} 
     \delta ( m_1 + m_2 + \cdots m_k - k + 1  , m )    \cr }}
The coefficients are matched with the 
 algebra by using a recursive procedure. 

Now we move on to the combination of the 
left and right operators which give the
full chiral primaries. 
Using the commutators of the chiral sectors 
of the form above, we can write down 
the commutators of the non-chiral operators
which contain terms written below
\eqn\ncom{\eqalign{ 
&  \sum_{ \sigma_{m_1} \in T_{m_1},  \sigma_{ m_2 } \in T_{n_2} } 
\bigl[ O^{\sigma_{m_1} }  {\bar O}^{ \sigma_{m_1} },  
           {\bar O}^{ \dagger \sigma_{m_1}}  {\bar O}^{\dagger \sigma_{m_2} }
     \bigr] =  \cr  
& \sum_{k} \sum_{\sigma_{3} \in T_{n_1,n_2 \cdots n_k} } N^{-{ k\over 2}  } 
  \delta ( m_2 -m_1 + k , n_1 + n_2 + \cdots n_k ) \cr 
& \qquad \qquad \qquad
   O^{ \dagger \sigma_{n_1}} 
     O^{\dagger \sigma_{n_2}}  \cdots O^{\dagger \sigma_{n_k} }
{ \bar  O}^{ \dagger \sigma_{n_1}} 
     { \bar O}^{\dagger\sigma_{n_2} }  \cdots {\bar O}^{\dagger \sigma_{n_k} }
 \cr }}
which have a similar structure to the terms in 
\chircomi.


A free field representation for
the algebra can be constructed where 
\eqn\alg{ A_{-n} = \alpha_{-n} + 
          { 1 \over \sqrt N } ( 1 + 
 { \cal O }(  { 1\over N} ) ) ( \alpha \alpha + \cdots  ) + \cdots 
         { 1 \over N^{ l/2 }  }  ( 1 + { \cal O} (1/N )  ) 
                 \alpha^{ l+1 }   }
 
 The second term in the above expression 
 is quadratic in the $ \alpha$'s. 
 The higher term  weighted by $ N^{-{ l  \over 2}}$ 
 is a polynomial in the $\alpha , \alpha^{\dagger} $ of degree 
 $ l+ 1 $. It is noteworthy that, for fixed degree of the polynomial, 
 the expansion involves powers of $ 1/N$ and not 
 ${ 1 \over \sqrt N}$. The odd powers of $\sqrt {N}$ can be 
 removed by redefining the operators.  
 This is  in agreement with the idea that the 
 algebra is closely related to the exclusion principle 
 and to quantum groups, which both  involve the parameter
 $N$, and not $\sqrt {N}$.

 There  are extra operators appearing on the 
 RHS of the equation have the form 
\eqn\frmext{ 
  O_{n} \bar  O_{m}( \bar  O_{n-m })^{\dagger}
  = \alpha_{n } \alpha_{-m} +  \cdots      }
The composite operators have also 
 been normalized to have unit two-point function. 
 They are subleading in the $1/N$ expansion. 

 \subsec{ Hamiltonian} 
The free field realizations  of the 
algebra in the large $N$ limit provide
representations of operators of the theory and of the Hamiltonian.
Based on studying the free field 
 reps of sub-algebras of the full algebra (in particular the
global case of the previous section) we expect 
 that there  exists first a free representation 
 where the Hamiltonian is quadratic : 
\eqn\repfr{ 
H^{(f)} = \sum_{m,p,q} ( m  + { p+q \over 2 } ) 
                 \alpha_{-m}^{p,q} \alpha_{m}^{p,q}   
}
In this picture  the interactions are contained in the nonlinear forms of
the other (in particular the creation-anihilation ) generators. 
We expect another representation
 where the Hamiltonian is nonlinear and exibits interactions in 
$1/N$ .For the  chiral primaries represented by $(0,0),(1,1)$ and $(2,2)$
forms  this takes the form 
\eqn\repi{\eqalign{& 
H^{I} = \sum_{m,p,q} \sum_{m,p,q} ( m  + { p+q \over 2 } ) 
                 \alpha_{-m}^{p,q} \alpha_{m}^{p,q}    \cr
&  + \sum_{n,k} { 1 \over \sqrt N }v(k,n) 
\bigl[ \alpha_{-n-k+1} \alpha_{k} \alpha_{n} 
         + \alpha_{-n} \alpha_{-k} \alpha_{n+k-1 }  \cr 
& +
(  \alpha_{-n-k+1 }^r \alpha^r_{k}\alpha_{n} + h.c. )   
  + ( \alpha_{-n-k+1}^{(2,2)} \alpha_{k}^{(2,2)}  \alpha_{n} + h.c. ) 
   \cr 
&  +  (  \alpha_{-n-k+1}^{(2,2)} \alpha_{k}^r \alpha_{n}^r  + 
  h.c )  \cr 
&  +         (  \alpha_{-n-k+3}^{(2,2)} \alpha_{k} \alpha_{n}  
       + h.c )  \bigr] \cr }}
with the form factor
\eqn\frmexti{
v({k},{n})={ \sqrt { (n+k)nk }}  } 
and with appropriate numerical coefficients . 
 The above form  would represent the leading term 
 with higher interactions in
 powers of $N$. Concerning this Hamiltonian and the emerging
 $1 +1$ dimensional field theory
one has the following comments. It effectively summarizes the dynamics
of chiral primaries with their correlation functions.
The dimension conjugate to the twist $n$ corresponds
to a coordinate  of AdS obtained after chiral primary reduction.
The structure of 
the 1+1 dimensional hamiltonian and its form factor $v(k,n)$ is identical
in form to the collective field theory \dj\ of 2d strings. 
The fact that there
is an analogy between the 2d noncritical string and radial dynamics in
the AdS/CFT correspondence has been observed earlier,for example
\refs{\gkp,\vp }.
We would also like to stress that the dynamics outlined above 
is universal. Chiral primaries in other AdS theories ($ AdS_5 \times S^5$
or $AdS_4 \times S^7$ ) are also described by an analogous one plus 
one dimensional hamiltonian. This follows by the fact that 
the form factor $v(k,n)$
seems to be the same in all these theories. In the present description
it reflects the fact that behind the AdS/CFT correspondence there
is an underlying non-critical string theory describable by
coll. field theory.

The above structure (of the algebra) 
and the hamiltonian is operational on the space 
of chiral primary operators. We should mention
the  extension to a more general
class of states as follows: using the SUSY algebra 
would provide couplings to
other fields of the full short multiplet, likewise the lowering
 operations of both
$SL(2)\times SL(2)$ and $SU(2) \times SU(2)$ would specify couplings to the 
corresponding
descendants spanning the full $AdS_3 \times  S^3$ space-time.
At leading orther
this extensions are direct, 
at higher order in $N$ they remain  a 
challenge. The non-linear realizations of some of the 
 basic symmetries \jky\  might play an important role in these
extensions.

\newsec{  Exclusion Principle and the Lie Algebra of observables. } 
 
 The properties of the Lie algebra associated with the 
 chiral primaries are closely related to the 
 Exclusion principle. 
 To clarify this we will need to review a few facts 
 about the chiral ring of the $S^N(X)$ theory,
 all of them related to the finiteness properties,  
 which are referred to as the exclusion principle. 
 
The simplest fact under this heading is that 
 there an arbitrary element of the ring 
 generated by the $A$ operators of section 2 
 has its left and right $SU(2)$ quantum numbers bounded
 as  $ 2j_L \le 2N $ and $2j_R \le 2N$. This follows 
 from the unitarity constraints of the $N=2$ superconformal sub-algebras
of the $N=(4,4)$ symmetry 
 \lvw. For example $ ( A_{-2}^{(0,0)} )^{ 2N + 1 } $ necessarily 
 vanishes by this argument. 
 Another aspect of the exclusion principle, emphasized in I,
 refers to the properties of single particle states, which 
 can be defined in the CFT as a linearly 
 independent set of  
 chiral primary operators which cannot be written 
 entirely in terms of products of other chiral 
 primaries. They are the analogs of single trace operators 
 in   Yang Mills. It was found in I that these generators
 are also cutoff by their $SU_L(2)$ and $SU_R(2)$
 quantum numbers, and this was an important piece of evidence 
 in favour of a non-commutative spacetime. 
 Another property of the vector space of chiral primaries
 is that it has a basis defined in terms of Fock 
 space creation operators $ B_{-n}^{\omega^{p,q} } $, which are related
 but not identical to the $A_{-n}^{\omega^{p,q} }$. 
 We can describe a vector space $H_{B}$ 
 as a truncated Fock space. Its states are in 1-1 correspondence 
 with 
\eqn\smq{
 B_{-n_1}^{\omega_1}  B_{-n_2}^{\omega_2} \cdots B_{-{n_k}}^{\omega_{n_k}} |0> 
} 
 with the restriction 
 $\sum_{n_i} n_i = N $. 
 The relation between the space of chiral primaries
 defined as generated by $A_{-n}^{\omega}$
 subject to relations between them which follow from 
 the OPE's is as follows. Let these relations 
take the form 
\eqn\rels{
R_i( A  )  = 0 
} 
 COnsider the ring of polynomials in the 
 $A_{-n}^{\omega}$ quotiented by the relations 
 above, and let this space be $H/R$ where 
 $R$ is the ideal generated by the 
 $R_i$. We have an isomorphism between $H/R$ and $H_B$. 
 
The relations $R_i$ are closely related to the structure
 of the Lie algebra we constructed from the 
 $A$ and $A^{\dagger}$. Setting these elements to zero 
 has to be consistent with the algebra. In other words 
\eqn\ide{ 
[ L_a , R_i ] = C_{aijb} L_b R_j 
} 
where $L_a$ and $L_b$ are elements in 
the Universal Enveloping algebra of the
Lie algebra. 
An example of this kind of relation 
in the case of a simple $sl(2)$ subalgebra
 as given in I. In fact the unitarity 
 conditions in the algebra were used  to 
{  \it derive }  these relations.

\subsec{ q-deformed super-algebra structure of fusion rules } 

For simplicity we restrict the discussion here 
 to couplings involving only the pure-twist operators 
$A_{-n}^{(0,0)}$. For 
 $ n \le N$ these are generating 
 elements of the chiral ring.
For small values of $n$ we have fusion of the 
 form 
\eqn\twsi{
A_{-n}^{(0,0)} A_{-m}^{(0,0)} \sim A_{-n-m+1}^{(0,0)}   
}  
There can also appear on the RHS products 
 of $A$'s but we can restrict  attention to 
  the terms on the RHS which are relevant for 
 studying the 3-point functions of generators of the chiral ring. 
On the left we have operators of $SU(2)_L$
spins $2j_1 = n -1 $ and $2j_2 = m-1$. On the right 
 we have $ 2j = 2j_1 + 2j_2 $. This is consistent  
 with $SU(2)$ fusion rules but not all $SU(2)$ reps. allowed
 by $SU(2)$ tensor products appear on the RHS. 
 Rather a good model for the fusion is given 
 by the tensor products of short reps. of 
 $SU_L(2|1,1)$ ( since the $SU_R(2|1,1)$ quantum numbers
 are identical to the left moving ones it suffices to focus 
 on the left moving symmetry). The spin $j$  then 
 labels the  largest $SU(2)$ spin  present in the decomposition of 
 the short $SU(2|1,1)$ rep. into reps. of its $SU(2)$ subalgebra. 

A qualitatively new feature appears 
when $n$ comes close to $N$, and $n + k \ge N$. 
 Then  
\eqn\bdt{
A_{-n}^{(0,0)} A_{-k}^{(0,0)} \sim 0   
}  
We can see this explicitly from the formulae 
 for correlation functions we wrote down in 
\thpf. We rewrite the equation relevant  
 for this  feature here : 
\eqn\bdtcor{   \langle O_{n+k-1}^{(0,0) \dagger}(\infty) 
O_{k}^{(0,0)}(1)\,O_{n}^{(0,0)}(0)\rangle =  
({{(N-n)!\,(N-k)!\, (n+k-1)^3 } \over 
{(N-(n+k-1))!\,N!\,n\,k}})^{{1 \over 2}   } } 
 We see that when $n+k-1$ exceeds $N$  
 the denominator contains a $\Gamma$ function 
 with a negative integer argument which causes 
 the expression to vanish. 
This feature can be explained  using 
fusion rules of $SU_q(2|1,1)$ where $q$ is, 
as in I, given by $q = e^{i \pi \over N+1}$. 
The q-deformed superalgebra will have a family
 of  unitary short ( atypical ) reps. which will 
 include unitary irreps. of the $SU_q(2)$. 
 Since all the reps. of the $SU_q(2)$ appearing 
 in the decomposition from $SU_q(2|1,1)$ to 
 $SU_q(2)$ have to be unitary, there will be a bound on these 
 short reps.  $2j \le N -1 $. 

\subsec{ $SU_q(2|1,1)$ and multiparticle states } 
 We have explained the qualitative features
 of the couplings between this family of generators 
 of the chiral ring, using $SU_q(2|1,1)$. 
 It will be very interesting to find a more detailed
 comparison between the fusions of the chiral ring 
 and the q-deformed super-algebra. 
 Some novel features, unfamiliar from ordinary rational 
 CFT, but that have appeared in studies of 
 WZW on supergroups \rsal\  have to be taken into account. 
 For example it has been found that one needs,  in general,  
 to take into account the indecomposable reps. of the superalgebra
 as well. It is very plausible, that in this case, we also 
 need to include reps. of $SU_q(2|1,1)$ which contain 
 indecomposables of the $SU_q(2)$ sub-algebra.  
 While the $SU_q(2|1,1)$ with the $q$ given above, nicely 
 describes the cutoffs on generating $A_{-n}^{(p,q)}$ \jr, 
 it also seems to have enough structure to account for 
 some other cutoffs in this theory. For example, 
 $SU_q(2)$ has, in addition to the standard irreps, 
 with a cutoff  $2j \le N-1 $, a family of irreps 
 $I_z^{p}$ in the notation of \Keller .  
 The detailed form of these reps. is given 
 in \reshtu.  
 Usually one drops these reps. in studying 
 standard connections between $SU_q(2)$ and the 
 and $SU(2)$ WZW, but in this model the chiral primaries
 transform in more complicated reps. of the $SU(2)$ current 
 algebra than the standard integrable reps. \gepwit\mathint\ 
 usually considered in  $SU(2)$ WZW of level $k=N$.
  We may expect a larger class of $SU_q(2)$ reps. 
 to appear in the corresponding quantum group model. 
 
 We mention a  very suggestive numerical observation 
 in favour of the above line of argument. 
 The family  $I_0^{p}$ has a cutoff at $ 2j \le 2N$. 
  The $A_{-n}$ operators coming from twisted sectors
  can be raised to powers which allow them 
 to exceed the bound at $N-1$.  
 It turns out, however that that there 
 is  a cutoff on the highest weight $2j \le 2N$ which works 
 for any chiral primary, as explained in the previous section. 
 This means that if we associate reps containing these  
 indecomposables to some of the chiral primaries which are products 
 of $A_{-n}^{p,q}$'s, we can explain both the cutoff 
 $2j \le N-1$ for the generators, and the cutoff 
 $2j \le 2N $ for arbitrary chiral primaries.

\subsec{ Black Holes and Correlation functions 
 of Chiral primaries } 
We saw that the Exclusion Principe
 is one manifestation 
 of the breakdown of the free-field
 representation of the algebra we have been 
 considering. 
This algebra, as we have emphasized, is closely related 
to the correlation functions. 
 We saw at the beginning of this section 
 in the discussion of \bdt\ and \bdtcor\ 
 that the exclusion principle 
 shows up in a qualitative change in behaviour of the
 correlation functions as a function of a parameter appraoching $N$. 
 We also saw in the discussion of the untwisted 
 sector algebra ( see in particular  \dervan ) 
 that correlators with a large number of 
 insertions start to show non-free field behaviour. 

 We would like to use the  properties
 of the orbifold CFT to find the lowest conformal 
  weight  where we may expect a divergence from free field 
 behaviour. 
 Recall the discussion of the map to free 
 fields. 
 We had for example 
\eqn\ffm{\eqalign{&  A_{-2}^{(0,0)} = \alpha_{-2} + \cdots \cr 
                 & A_{-3}^{(0,0)} = \alpha_{-3} + \cdots   \cr }}
However we do not have
\eqn\ffmi{ 
( A_{-2}^{(0,0)})^2 = \alpha_{-2}^2 + \cdots } 
This would contradict the fact that 
the correlation function 
 $ \langle A_{3} A_{-2} A_{-3} \rangle $ is non-zero at 
 order $ { 1 \over \sqrt N }  $. Rather the object 
 which behaves more like $ \alpha_{-2}^2$ is 
 the twist operator associated with the conjugacy 
 class with two cycles of length $2$, which is one of the terms 
 in \ncom\ for example.  
 While $A_{-2}^{l} $ is not, for small $l$,  
 a free oscillator raised to the $l$'th power,
 its behaviour in correlation functions 
 should not be qualitatively different from a power
 of a free field because it is a linear combination 
 including an operator which behaves like $ \alpha_{-2}^l$. 
 When $l$ hits $N/2$ the corresponding operator ceases
 to exist because we cannot have a permutation with 
 more than ${ N \over 2 }$ cycles of length $2$. 
 We observe that this happens at $L_0 ={  N \over 4}$. 
 If we try the same thing with 
 $A_{-3}^{l}$ we get a threshold of $l = { N \over 3} $ 
where we may expect a qualitative change in behaviour. 
 If we consider a general 
 operator of the form 
\eqn\gentr{ 
 ( A_{-2} )^{n_2} ( A_{-3} )^{n_3} \cdots A_{-n_k}^{k} } 
the corresponding free operator ceases 
 to exist when 
$ 2n_2 + 3 n_3 + \cdots  kn_k = N$. 
It has 
\eqn\wegt{\eqalign{   
 L_0 &=   { 1 \over 2 } ( n_2 + 2n_3 + \cdots k n_k ) \cr
     &= { 1 \over 2 } ( N - (n_2 + n_3 + \cdots n_k ) \cr
     &=  { 1 \over 2 } ( { N\over 2} +  
    { 1 \over 2 } n_3  +  { 2 \over 2 } n_3 + \cdots    + 
                                       { ( k -1)  \over 2 } n_k) \cr       }}
It is now clear that the lowest threshold 
 we get is ${ N \over 4 }$. 
Considering operators of type $A_{-n}^{p,q}$
 with $p,q \ne 0 $ only increases the threshold. 
Precisely this value  $L_0 = N/4 $ was obtained 
by \db\mms\ as the threshold where black holes 
 start to be relevant.  
We have argued that the same threshold 
 appears if we ask for the lowest value of $L_0$ where 
 operators in the chiral ring start to display 
 behaviour qualitatively different from free fields. 
 It will be very interesting to characterize
 the corresponding change of behaviour in the correlation functions
 and compare with expectations from black hole physics.

\newsec{ Conclusions } 

We have studied in this paper
a Lie algebra associated with 
the chiral primaries of $S^N(T^4)$ CFT 
and their CFT conjugates. 
The structure constants of this Lie algebra 
are simply related to the correlation functions
after an appropriate choice of basis. 
We obtained by CFT computations some of these structure 
constants, those which are determined by the 
 3-point functions 
of bosonic chiral primaries. 
This leads to a dynamics described by
an effective 1+1 dimensional
field theory which corresponds to 
 the simplest representation of the algebra.

Connections between the Lie algebra 
and the stringy exclusion principle 
were studied in I, 
 where the relation  with q-algebras and 
non-commutative spacetimes was emphasized.
These aspects  were elaborated 5.1-5.3. 
In 5.4 we focused on a characterization 
 of the exclusion principle as a
 deviation of the chiral primaries 
from free field behaviour in correlation functions, 
 and we were lead to look for the lowest threshold where 
such  deviation is expected. We found that the lowest 
threshold is at $L_0 = {N \over 4}$ which has been 
  argued to be relevant to black holes in $ADS_3$.
Developing further the relation between these 
 correlation functions and black holes will be very interesting.

The Lie algebra is a much richer 
structure than the truncated Fock space of
the chiral primaries themselves. 
The latter is only one representation of the Lie 
 algebra. This simplest representation was studied 
 in connection with actions associated to the 
 correlation functions of chiral primaries. 
 This was done in detail in section 3 
for the untwisted sector. A coherent state basis 
 was useful there, and we explained how the 
parameters in the coherent states are related to 
the multi-oscillator Heisenberg algebra at large $N$. 
In section 4, we pursued the study of the 
 large $N$ map between the Lie algebra and 
the multi-oscillator Heisenberg algebra ( free fields ).

Our discussion naturally raises 
 the question of other representations 
of the algebra which are not directly 
 related to the Fock space of the chiral primaries.     
These  will be important in 
studying the stringy states, which are certainly  
 important since the free orbifold CFT 
 is expected to be dual to a gravitational background
 where the graviton as well as other stringy 
 degrees of freedom are important,
 as emphasized in \refs{  \bhdm, \ml, \sw }.
 Studying the stringy states as representations 
 of this algebra derived from chiral primaries
 and their conjugates is particularly 
interesting, since the chiral primaries 
 and their truncations  
 show clear evidence of a non-commutative spacetime. 
This point of view on the stringy states should
 clarify the relevance of the non-commutative  
 space-time to the dynamics of the full set of stringy degrees 
 of freedom.

 The precise role of the non-commutative 
 spacetime in determining the dynamics
 of this gravitational theory remains to be further studied,
  but the simplest possibility is that, 
 in analogy to the case of non-commutative Yang Mills \cds\ 
 there is `simple action' on a non-commutative 
 spacetime which is related to a `complicated action' 
 on a commutative spacetime. The fact that 
 the techniques of collective field theory 
 allow the construction of spacetime actions starting from the CFT 
 should allow one to explore the above possibility.
 Other approaches for  studying the non-commutativity 
 of spacetime coordinates in the context of string theory 
 on  ADS-type backgrounds have been suggested in 
 \tesh\chk.

\bigskip

 \noindent{\bf Acknowledgements:}
 We are happy to acknowledge enjoyable and 
 instructive discussions with Miriam Cvetic, Sumit Das,  Pei Ming Ho, 
  Robert de Mello Koch, Vipul Periwal, Radu Tatar, 
 T. Yoneya, E. Witten. 
 M.M. would like to thank the organizers of TASI99
 for hospitality while part of the work was being done.  
 This research was supported by 
 DOE grant  DE-FG02/19ER40688-(Task A). 

\bigskip
\newsec{Appendix}
Let us consider first the OPE we used which can be read from 
\Frol where
it is detemined by a geometric construction. We will also 
neglect the fact that in our case the theory is a $T^{4}$ orbifold,
whereas in 
\Frol, it was a noncompact orbifold. The result for the constant is taken 
from equation (5.24) in \Frol with the following redefinitions:
\eqn\red {
N=n+1, ~~ n_{0}=n, ~~ D=6, ~~ p_{1}=k_{n}, 
~~ p_{2}=0, ~~ p=k_{2};
}
 \eqn\strc {
C(2,n|n+1;k_{2},k_{n})=2^{a}\,n^{b}\,(n+1)^{c},
}
where:

\eqn\strconst { \eqalign{ & a=-{9  \over 8}+
{k_{2}^{2}  \over 4},\cr  & 
b=-{3  \over 8}-{1 \over {n+1}}~({1  \over 4}~n^{2}-
{1 \over 2} k_{n}^{2} -~k_{2}~k_{n} - 
{{n-1} \over 4}~k_{2}^{2} ),\cr  & 
c={1 \over 8}+{1 \over 4}~(n+{1 \over n})-
{k_{2}^{2}  \over 4} -{k_{n}^{2}  \over {2~n}} . \cr
 }}
\par
We outline here the derivation for the equations used in
 the text only for the twist field which does not involve any $T^{4}$
indices, 
$O_{(1..n)}(u,\bar{u})$. In 
this case, in the equations \strc we take 
$k_{2}=(0,0,0,0,{1 \over 2},{1 \over 2})$,
$k_{n}=(0,0,0,0,{{n-1} \over 2},{{n-1} \over 2})$
 which give $a=-1,~
b=-{1 \over 2},~c={1 \over 2}$ and we have the first 
equation in \opr. Combining this result with an identical
 one for the right movers we obtain:
\eqn\oprt{ O_{(n\,n+1)}(u,\bar{u})\,
O_{(1..n)}(0) =  {1 \over 4}\,{{n+1 } \over {n}}
 ( O_{(1..n+1)}(0) + 
O_{(1..n+1\, n)}(0))}
This result has a clear interpretation for the case
 when the lenght of the cycle at
$0$ is set to $1$, namely $n=1$ case.
 We used this extrapolation for the other cases
to arrive at equations \opera.
\par
Starting from equation \oprt\ we can use now the 
conformal field theory rules in order to determine more involved
correlators.
 We look to an OPE involving three twists:
\eqn\ind
{O_{(n+k-1\,n+k)}(u,\bar{u})~
O_{(n..n+k-1)}(z,\bar{z})~ 
O_{(1..n)}(0)=const.~(O_{(1..n+k)}(0)+...)}
where $...$ stand for the other 4 possible terms,
 and the right hand side is in the limit $u \rightarrow 0;~z \rightarrow
0$. 
For consistency, the limit should not depend on the 
order, and from this we derive the following 
recurrence equation for the constant which appear in the OPE of a twist of 
lenght $n$ and of one of lenght $k$, denoted in
this appendix as C(n,k):
\eqn\indu { \eqalign
{ & C(n,k+1)=C(n,k)~{{(n+k)~k} \over {(n+k-1)~(k+1)}}, \cr
  & C(n,2)={(n+1) \over {4~n}}. \cr
}}
These equations are solved by :
\eqn\sol { C(n,k)={{n+k-1} \over {2~n~k}}}
It is interesting to observe that in this derivation of the constants 
appearing in the 
OPE of the twist operators it was helpful to use both the geometric
construction
\Frol and the associativity of the OPE in conformal field theory.
\par
Using a similar derivation, we are able to obtain also the OPE 
for the twist operators  
overlapping over a two cycle.

\listrefs

\end